# Modeling Stakeholder-centric Value Chain of Data to Understand Data Exchange Ecosystem

Teruaki Hayashi[1][0000-0002-1806-5852], Gensei Ishimura[2][0000-0001-7091-1473], and Yukio Ohsawa[1][0000-0003-2943-2547]

[1] The University of Tokyo, 113-8656, 7-3-1, Hongo, Bunkyo-ku, Tokyo, Japan
[2] Professional University of Information and Management for Innovation, 131-0044, 1-18-13, Bunka, Sumida-ku, Tokyo, Japan
`hayashi@sys.t.u-tokyo.ac.jp`

**Abstract.** In recent years, the expectation that new businesses and economic value can be created by combining/exchanging data from different fields has risen. However, value creation by data exchange involves not only data, but also technologies and a variety of stakeholders that are integrated and in competition with one another. This makes the data exchange ecosystem a challenging subject to study. In this paper, we propose a model describing the stakeholder-centric value chain (SVC) of data by focusing on the relationships among stakeholders in data businesses and discussing creative ways to use them. The SVC model enables the analysis and understanding of the structural characteristics of the data exchange ecosystem. We identified stakeholders who carry potential risk, those who play central roles in the ecosystem, and the distribution of profit among them using business models collected by the SVC.

## 1  Introduction

Data have been treated as economic goods in recent years, and have begun to be exchanged and traded in the market [1, 2]. Interdisciplinary business collaborations in the data exchange ecosystem have been springing up around the world, and the transaction of data among businesses has garnered the interest of researchers [3, 4]. Moreover, the expectation that personal data can be valuable has risen [5, 6], and an increasing number of businesses are entering the market that forms the ecosystem of data exchange. However, value creation in the ecosystem involves not only data, but also technologies and a variety of stakeholders that are integrated and competing against one another. This makes the ecosystem a difficult subject of research. In other words, few frameworks are available to understand and share the functions and interactions of interdisciplinary data businesses. If personal data are leaked, for example, it is difficult to determine who is responsible, what caused it, or where the bottleneck of the data business is. The deposit of data and control rights as well as the transfer to third parties increases the complexity of stakeholders related to the data business, and makes it difficult to control the value chains. Therefore, establishing an appropriate unit of analysis and framework for a comprehensive understanding of the data exchange ecosystem is an important issue.

In this study, we propose a model for describing the stakeholder-centric value chain (SVC) of data by focusing on the relationships among stakeholders in the data businesses, and discuss creative ways to use these relationships. The contribution of this study to the literature is in describing interdisciplinary data businesses in a simple manner with minimal expression. We discuss this description and its functionality as a common language across data businesses, and the support systems that can be implemented by forming a knowledge base of data businesses collected using the SVC model.

## 2    Data Exchange Ecosystem and Relevant Studies

As represented by digital transformation, digitization and data collaboration are expected to become prevalent in society. Unlike the conventional supply chain, the business models of interdisciplinary data exchange have been decentralized—sharing roles and values in flat and even relationships among the stakeholders. In the business model manifested by the business of e-books, for example, the computer industry, home appliances, publishers, and telecommunication companies dynamically work together to form a complex ecosystem [7]. To propose technologies and understand the characteristics of the ecosystem, several studies have attempted to tackle these challenges.

As a technology for data exchange, the Innovators' Marketplace on Data Jackets (IMDJ) provides the framework for discussion among stakeholders in the data exchange ecosystem—data holders, users, and analysts—to lead data collaboration and innovation [8]. The Data Jacket Store structures and reuses knowledge for data utilization generated in the IMDJ [9] and analyzes the relationships [10]. However, the relevant studies do not discuss human relationships in the data exchange ecosystem. As representative of research on stakeholders, action planning focuses on their roles and proposes a descriptive model for data businesses [11]. The Industrial Value Chain Initiative (IVI) offers tools that support interdisciplinary data collaborations using the software IVI Modeler (https://iv-i.org/wp/en/), which has 16 types of charts to describe a business model. Deloitte LLP describes the relationships among stakeholders in the open-data marketplace by such roles as data enablers, suppliers, and individuals [12]. These studies provide frameworks for explaining and sharing business models, but do not discuss ways of understanding the ecosystem from a macroscopic perspective.

## 3    Stakeholder-centric Value Chain of Data

The SVC model is an approximate unit of analysis to understand business models in the ecosystem of data exchange that focuses on the stakeholders. In such applications of it as in knowledge representation, the SVC model is based on graph representation that uses nodes, edges, and labels. Table 1 describes the elements and labels of the SVC model in this context. Nodes have two types of attributes: individual and institution. Entities such as data and services exchanged among the stakeholders are defined as labels at the edges of the directed graph. An edge has six types of labels, and the data they represent are further divided into three types: a collection of non-personal data and personal data, and personal data of each individual. Many methods are available for data provision, for example, downloading data stored on a website or obtaining them by APIs such that their details are expressed in the relevant comments. There are various types of data processing, such as data cleansing, which in turn includes anonymization, visualization, and such AI techniques as machine learning. The details of the data processing are described in parentheses or in the comments. Note that the data processing is described in a self-loop in the SVC model. To consider the time-series information of data businesses, timesteps are attached to the edges as attributes, and if the nodes and edges require additional explanation, we can add comments to them. To describe data businesses, users can employ graphical icons (pictograms) to easily share and understand the overall structure of the business models. Yoshizawa et al. proposed a semantic network for knowledge-sharing using pictograms [13]. This method introduces a network of elements, including humans and actions, and is intended to be used in education, which is a different domain from the one considered here.

A combination of the relationships between stakeholders—the smallest unit of knowledge—allows for the description of stakeholder-centric data businesses, and the integration of these business models helps structure the ecosystem of data exchange. We use a network-based approach to analyze the characteristics of the data exchange ecosystem. To efficiently encode data businesses described by the SVC through these networks, the $n$-th data business $G_n$ is represented as $(V_n, E_n, A_n, T_n)$, which is a directed multi-attribute graph. $G_n$ consists of nodes $v \in V_n$, a set of stakeholders, edges $e_{ij} \in E_n$, and a set of relationships between the $j$-th and the $i$-th nodes. In this model, the values of the attribute are not numerical but given as a set of labels, where $A_n$ represents two attribute values of the node, and $T_n$ is a set of six attribute values of the edges (Table 1). Each node has only one attribute, whereas each edge can have an unlimited number of attribute values depending on the number of its relationships.

**Table 1.** Descriptive items of the SVC model.

| Element | Label | Icon | Description |
|---|---|---|---|
| Node | Individual |  | A label used when the stakeholder is an individual. |
| | Institution |  | A label used when the stakeholder is a company, an institution, or a group of individuals. |
| Edge | Request | R | A label indicating a request for the data. |
| | Service | S | A label indicating the service offered including the product. |
| | Payment | $ | A label that indicates the exchange of money (payment). |
| | Data | 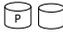 | A label that represents the data provided. "P" represents personal data, and subscript $i$ identifies an individual if needed. |
| | Process | Proc() | A label that represents data processing. The name of the algorithm can be described in parentheses. |
| | Timestep |  | The time order is represented by the subscript $j$. If a branch occurs, hyphens or underbars can be used. |
| Other | Comment |  | Additional information about nodes and edges. |

In light of the discussion above, users of the SVC model and the described data businesses can be (A) partners of the data business, or (B) those who are not involved in the data business. The benefits for (A) are that the stakeholders can clarify the purposes of the business and confirm that there are no misperceptions or inconsistencies among the business partners. They can also explain the essence of the business to potential partners and look for new collaborations. By contrast, for (B), the users can understand the purpose and conditions of data businesses as well as how the stakeholders participate in their data businesses. They can also analyze the applicability of the business models and potential partnerships, thereby increasing their business values.

## 4    Experimental Details

One hundred-and-five participants were involved in our experiments. They consisted of engineering students and business people over 20 years of age who were interested or engaged in data businesses. We allowed them to form groups of two or three, and lectured them on the SVC model for 30 minutes. We then asked each group to describe the outline of data businesses using the SVC model in 30 minutes. This yielded 35 data business diagrams. We focused on a simple description of a business model, and did not rigorously evaluate the completeness and accuracy of the described business models. Therefore, note that business models do not always represent the empirical relationships among businesses.

## 5    Results and Discussion

### 5.1    Structural Characteristics of the Data Exchange Ecosystem

Figure 1 shows the relationship diagram that integrates 35 business models with 123 stakeholders as obtained in the experiment. In the diagram, companies/institutions are represented by square nodes and individuals by circular nodes. The size of each node represents the frequency of the relevant stakeholders. The labels are listed by edges, and thickness represents the number of labels. The largest business model in terms of stakeholders had 10, and the smallest had two. The average number of stakeholders in each business model was 4.97.

The network of all business models was divided into five subgraphs, and we analyzed the largest component (the number of nodes: $|V| = 93$; the number of edges: $|E| = 128$). Note that self-loops were not included, and the values were calculated as an undirected graph. From a macroscopic point of view, the average degree ($\langle k \rangle$) was low at 2.75, and the power index $\gamma$ was 2.44, where this represented the power distribution. The density was very small at $\rho = 0.00299$, and the stakeholder network was globally sparse. Furthermore, the value of the clustering coefficient was $\langle C \rangle = 0.286$, and reflected the existence of hubs in local clusters. As shown in Fig. 1, the network consisted of many low-frequency stakeholders with many densely connected clusters. Data processors, data accumulators, and users appeared frequently across the business models, and were located at central positions in the network. It is also noteworthy that, unlike human-related networks in general as networks of co-author or actor relationships [14], assortativity [15] did not have a high positive value at $r = -0.128$. That is, the stakeholder network of the data businesses was disassortative or neutral, and had similar characteristics to those of engineering or natural networks (representing power grids or protein interactions, respectively [14]). In these networks, nodes with a high degree tend to have linkages with low-degree nodes. In other words, hub nodes in the network are connected to avoid other hubs. In the economic system, the data providers may not sell the data to other data providers, and data analysts may have a rivalry with one another, such that the business relationships among those who have the same roles in the market are unlikely to be cultivated. Such differences may lead to a segregation of stakeholders in the ecosystem. By contrast, hubs are not connected to other hubs, as in the case of social networks [14], and the redundancy of the network is low due to its sparseness. In other words, the stakeholder network easily disintegrates when the stakeholders are removed. Whether this feature occurs because the data exchange ecosystem is not yet a mature market is debatable. To improve the understanding of the mechanism, it is necessary to compare it with the stakeholder relationships in other markets, and test with more samples of business models and the details of their business processes [16].

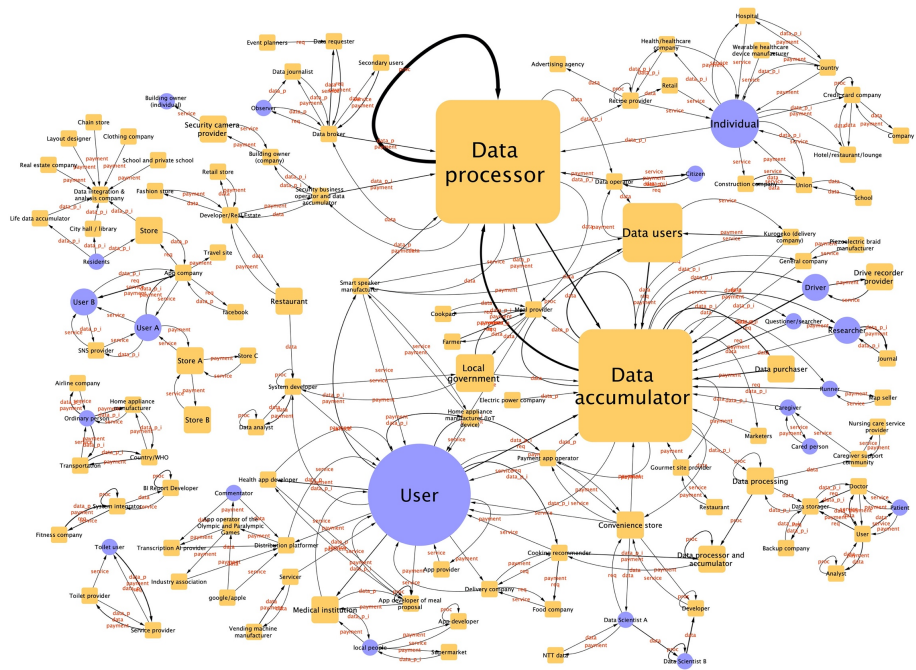

**Fig. 1.** The diagram of business models using the SVC.

## 5.2 Knowledge Extraction from SVC

The first purpose of the SVC model as used here was to understand the structural characteristics of the data exchange ecosystem, and the second was to reuse the structured knowledge of the data businesses. Because the knowledge unit of the SVC model had a structure identical to that of linked data, such as the RDF triple, it worked as a knowledge base for a search system. Below is information that can be gleaned from structured knowledge.

    (a) The most profitable stakeholder.
    (b) The stakeholder who carries potential risk, for example, if personal data are leaked.
    (c) The stakeholder who plays a central role in the ecosystem.
    (d) Profit distribution among stakeholders.

Functioning as an ecosystem means that payments are made appropriately for the provision of data and services. Although the SVC does not include the amount of payment, it was possible to identify stakeholders who get the most money across business models. As a result, we found that data processors were paid 18 times; data accumulators: 12 times; and data integration and analysis company: 5 times. Meanwhile, let us analyze who has the highest risk in the case of leakage of personal data. We can obtain the stakeholders by those who are the end points of the flows of personal data (data($p$) or data($p_i$) in Fig. 1). As a result, we obtained data processors got personal data 27 times, data accumulators: 13 times; and app company: 5 times. These results suggest that the data processor was most likely to receive the highest profit, but there was a potential risk of personal data leakage for it. Moreover, using the degree and betweenness centralities, we found that data users, accumulators, and processors were the top three stakeholders that appeared frequently across business models, and may play a central role in the ecosystem.

It is also possible to create an indicator to measure whether the services provided, such as data and products, are appropriately paid for in each business model or ecosystem. We defined $k_i^{\text{in}} = \sum_{j=1}^{|V|} |e_{ij}|$ to represent the in-degree of the $i$-th stakeholder, $k_i^{\text{out}} = \sum_{j=1}^{|V|} |e_{ji}|$ as its out-degree, and the payment to the $i$-th stakeholder as $k_i^{\text{in}}(\text{payment}) = \sum_{j=1}^{|V|} |e_{ij}(\text{payment})|$. The request for data was different from that for service, and we excluded $k_i^{\text{out}}$ from this. We also included data processing with a self-loop because this is necessary for business models. We defined the received profit sufficiency (RPS) as the degree to which the $i$-th stakeholder had been sufficiently paid for the service it provided, as in Eq. (1):

$$\text{RPS}_i = k_i^{\text{in}}(\text{payment}) / (k_i^{\text{out}} - k_i^{\text{out}}(\text{request})) \quad (1)$$

The calculated RPS value of each stakeholder, and RPSs of the medical institutions, hospitals, and supermarkets was one, indicating sufficient payments for the data and services that they had provided. By contrast, the RPS of the data processor was 0.75 and that of the data accumulator was 0.30. This indicates payments to the data processors and accumulators were not adequate, although they carried the risk of personal data leakage. The RPSs of residents and restaurants who were data generators were zero, which appeared to be necessary to correct the business models. We do not discuss the accuracy of the RPS here, but its concept is important in assessing the soundness of the ecosystem. The RPS can be calculated not only for each stakeholder, but also for each business model and the entire ecosystem. The RPS of the entire ecosystem using $\sum_{i=1}^{|V|} \text{RPS}_i$ was 0.32. To improve the soundness of the ecosystem, it is useful to review the business models and stakeholder relationships so that the RPS approaches one.

### 5.3 Limitations and Future Work

In the experiments, all participants were able to describe the model diagrams of the data businesses in a short time using the SVC model. One participant commented that the SVC model worked as a common language to explain and share the outlines of business models to the other members, suggesting that one of the purposes of our study was achieved. On the contrary, owing to an increase in the number of stakeholders and their relationships, the description of timesteps became complicated, and some businesses missed the information related to timesteps. Preventing excessive increases in the number of edges and excessive complexity of the timesteps are challenges for future research. Moreover, in this study, the edges had only labels, and we did not define their capacity to avoid complexity. By providing information on the frequency of data updates, task time, and flow rate at the edges, it is possible to apply methods used to solve maximum flow problems. Moreover, knowledge representation is generally based on the assumption that experts can describe the knowledge accurately and comprehensively. To understand all structures of the data exchange ecosystem, we must consider tacit knowledge in future research.

The method of analysis used here is simple but powerful in obtaining useful knowledge by applying network algorithms. Moreover, as the number of samples was small, we could not use other useful measures designed for complex networks, such as clustering or community extraction that considers multiple attributes along the edges and on nodes [17, 18]. In future research, it is important to collect more samples and verify the claims of this study.

## 6 Conclusion

The motivation of our study was to understand the ecosystem of data exchange and develop interdisciplinary data businesses. We proposed a framework for analyzing and understanding the structural characteristics of a data exchange ecosystem using a simple description model that focuses on the relationships among the stakeholders using a network-based approach. In spite of the promise of interdisciplinary data collaboration, due to the diversity of stakeholders and their complex relationships, it is challenging to understand this structure. Owing to its simplicity, the SVC model has the potential to function as a common language to understand data businesses. Moreover, when the number of stakeholders increases, the number of relationships exponentially increases as well. For this reason, it is difficult to consider every possible combination of stakeholders across areas. We think that SVC models and the reutilization of structured knowledge can support the development of the ecosystem as well as data businesses.

**Acknowledgement.**

This study was supported by the JSPS KAKENHI (JP19H05577 and JP20H02384), and the Artificial Intelligence Research Promotion Foundation.